# Feature Exploration for Knowledge-guided and Data-driven Approach Based Cuffless Blood Pressure Measurement

Xiaorong Ding, *Member, IEEE*, Bryan P Yan, Yuan-Ting Zhang, *Fellow, IEEE*, Jing Liu, Peng Su and Ni Zhao

*Abstract*—This letter explores extended feature space that is indicative of blood pressure (BP) changes for better estimation of continuous BP in an unobtrusive way. A total of 222 features were extracted from noninvasively acquired electrocardiogram (ECG) and photoplethysmogram (PPG) signals with the subject undergoing coronary angiography and/or percutaneous coronary intervention, during which intra-arterial BP was recorded simultaneously with the subject at rest and while administering drugs to induce BP variations. The association between the extracted features and the BP components, i.e. systolic BP (SBP), diastolic BP (DBP), mean BP (MBP), and pulse pressure (PP) were analyzed and evaluated in terms of correlation coefficient, cross sample entropy, and mutual information, respectively. Results show that the most relevant indicator for both SBP and MBP is the pulse full width at half maximum, and for DBP and PP, the amplitude between the peak of the first derivative of PPG (dPPG) to the valley of the second derivative of PPG (sdPPG) and the time interval between the peak of R wave and the sdPPG, respectively. As potential inputs to either the knowledge-guided model or data-driven method for cuffless BP calibration, the proposed expanded features are expected to improve the estimation accuracy of cuffless BP.

*Index Terms*—Extended feature space, cuffless blood pressure, full width with half maximum, pulse transit time.

## I. INTRODUCTION

CUFFLESS blood pressure (BP) measuring technique has attracted considerable attention with the advanced development of emerging sensing technology and pervasive computational capability, as well as the demand of monitoring hypertension outcomes and cardiovascular status in a continuous and unobtrusive way. As an alternative of conventional intrusive approach or cuff-based obtrusive techniques, cuffless method provides an indirect estimate of BP through modeling the relationship between BP and the cardiovascular indicators that can track the changes in BP. These indicators can be obtained by noninvasive and wearable sensors. The best known example of such indicators is the pulse transit time (PTT), which is the time it taken for an arterial pulse propagating from the aorta to a peripheral site. PTT can be easily accessed by means of two pulses originating from the cardiovascular system, such as the electrocardiogram (ECG) and photoplethysmogram (PPG). Principally as a measure of arterial stiffness, PTT can track variations in BP. Ever since the 1980s, many studies have conducted to explore the possibility of PTT to track BP changes. Following on that, many efforts have been devoted to investigating the usage of PTT to estimate BP such as to avoid the traditional cuff-based method. However, the fact is that PTT alone is insufficient to mirror all kinds of variations of dynamic BP. Although lots of works have focused on the calibration of PTT to BP, the problem of estimation accuracy still remains a challenge [1].

To address this concern, recent research efforts have been extended on two dimensions: the exploration of more efficient BP indicators/features [2, 3], and the modeling of the relationship between the indicators/features and BP [4, 5]. With the emerging application of machine learning (ML) based predictive algorithms, there has been increasing attempts using ML algorithm to improve the accuracy of BP prediction. Compared with the traditional regression models and explicit analytic models, ML is far more powerful to capture complex nonlinear relationships between input and output features [3, 5, 6]. For example, a recent study by Su et al used deep recurrent neural network to develop the BP model with 7 features extracted from ECG and PPG, which outperformed traditional PTT-based and regression method for both short-term and long-term BP monitoring [7]. The art of ML starts with the design of appropriate data representations, and better performance is often achieved with better features derived from the original input. However, for cuffless BP, current studies either have employed very limited handcrafted features or just used raw signals as inputs, which cannot result in a precise target.

In this study, we explore the features based on the linear and nonlinear aspects of the relationship between the extracted potential indicators and BP, with the identification of the best

*This project was supported by the Innovation and Technology Fund (Ref. No. ITS/275/15FP) from the Innovation and Technology Commission of Hong Kong and the TIM Seed Grant from the Chow Yuk Ho Technology Centre for Innovative Medicine. * indicates corresponding author.

X. R. Ding, J. Liu, P. Su and N. Zhao, are with the Department of Electronic Engineering, The Chinese University of Hong Kong, Hong Kong (e-mail: xiaorong.ding@eng.ox.ac.uk; jingliu@cuhk.edu.hk; psu@cuhk.edu.hk; nzhao@ee.cuhk.edu.hk).

B. P. Yan is with the Department of Medicine and Therapeutics, The Chinese University of Hong Kong, Hong Kong, China (e-mail: bryan.yan@cuhk.edu.hk).

Y. T. Zhang is with the Department of Biomedical Engineering, City University of Hong Kong, Hong Kong, China (e-mail: yt.zhang@cityu.edu.hk).

correlated or coupled features based on correlation coefficient, cross sample entropy and mutual information. This study will provide insight into the causal inferences between the indicators and BP, which is crucial for either knowledge-guided or data-driven method for cuffless BP estimation.

## II. METHODOLOGY

### A. Clinical Experiment

Continuous intra-arterial BP, ECG and PPG signals were collected with a sampling rate of 1000 Hz from two patients undergoing coronary angiography and/or percutaneous coronary intervention at Prince of Wales Hospital of The Chinese University of Hong Kong -. The protocol consists of signal acquisition with BP catheter putting in the central aortic for 3 min, followed by 2 min recording after administering intra-arterial nitroglycerine (50-200 micrograms) to induce transient BP changes, and finally moving the catheter to the radial artery and recording signals for 3 min. ECG and PPG were recorded with one-lead ECG electrodes, and PPG sensor on the index finger of the contralateral arm of the intra-arterial BP, respectively. The experiment has been approved by Joint Chinese University of Hong Kong – New Territories East Cluster Clinical Research Ethics Committee, and all subjects gave their informed consents prior to the experiment.

### B. Feature Extraction

A total of 222 features were extracted from ECG and PPG, mainly the latter, including the PTT, time duration (TD), pulse width (PW), amplitude (AM), intensity of PPG (PI), the amplitude (AM) of the first derivative of PPG (dPPG) and second derivative PPG (sdPPG) at fiducial points, area under the PPG curve (AR), as well as physiological meaningful relative index (RI) such as the relative rising time, dicrotic vertical position, dicrotic diastolic ratio, slope transit time [8], and so on. The fiducial points and the major features are depicted in Fig. 1.

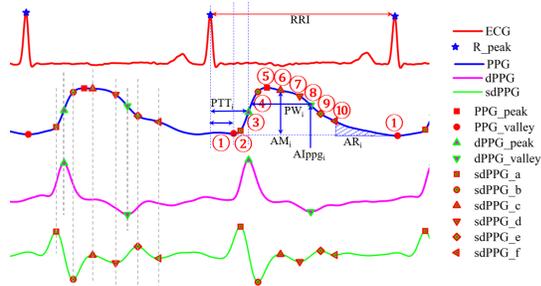

Fig. 1. Illustration of the fiducial points of electrocardiogram (ECG) and photoplethysmogram (PPG), as well as the main types of features.

The definition of the features was described as below as well as listed in Table I. All features were calculated per cardiac cycle:

**Feature point (FP$_i$, $i$=1~10)** = (PPG_valley, sdPPG_a, dPPG_peak, sdPPG_b, PPG_peak, sdPPG_c, sdPPG_d, dPPG_valley, sdPPG_e, sdPPG_f, PPG_valley_next)

**Pulse Transit Time (PTT$_i$)** = FP($i$) – R_peak, $i$ = 1~10

**Time Duration (TD$_i$)** = [RRI, (FP($j$) – FP($i$), $i$ = 1~10, $j \in i$, and $j>i$)]

**Pulse Width (PW$_i$)** = horizontal pulse width at specific level, *i.e.* the level at each fiducial point and at 50%, 60% and 70% of the pulse amplitude

**Amplitude (AM$_i$)** = PPG(FP($j$)) - PPG(FP($i$)), $i$ = 1~10, $j \in i$, and $j>i$

**Absolute Intensity of PPG (AIPPG$_i$)** = PPG(FP($i$)), $i$ = 1, 2, …, 10

**Absolute Intensity of dPPG (AIdPPG$_i$)** = dPPG(FP($i$))

**Absolute Intensity of sdPPG (AIsdPPG$_i$)** = sdPPG(FP($i$)), $i$ = 2, 4, 7-10

**Area under PPG curve (AR$_i$)** = Area between (FP($j$), FP($i$)), $i$ = 1~11, $j \in i$, and $j>i$

**Physiologically meant Ratio Index (RI$_i$)**: relative rising time, dicrotic diastolic ratio, augmentation index, inflection point area ratio, slope transit time, ratio of sdPPG ($b/a$, $c/a$, $(c+d-b)/a$, etc.), PPG intensity ratio, perfusion index [9].

TABLE I. Feature Index

| Index | Feature Category | Calculation Method |
|---|---|---|
| 1-10 | Pulse Transit Time (PTT) | Time difference between R peak of ECG and fiducial points of PPG |
| 11-66 | Time Duration (TD) | Time difference between fiducial points of PPG |
| 67-76 | Pulse Width (PW) | Pulse width at 50%, 60%, 70% and fiducial points of the PPG amplitude |
| 77-131 | Amplitude (AM) | Amplitude between fiducial points of PPG |
| 132-150 | Pulse Intensity (PI) | Intensity of PPG, dPPG, sdPPG at fiducial points |
| 151-204 | Area (AR) | Area under the PPG curve between fiducial points |
| 205-222 | Relative Index (RI) | Physiological meaningful ratio index |

### C. Feature Evaluation

The relationship between the extracted features and BP components, i.e. systolic BP (SBP), diastolic BP (DBP), mean BP (MBP), and pulse pressure (PP) were evaluated in terms of correlation coefficients. Since the correlation criteria can only detect linear dependencies between variable and target [10], we further investigate their associations with the cross sample entropy and mutual information.

The cross sample entropy provides an indication of the degree of synchronizing between two time series, the higher the entropy value, the less the synchronization between these two series, and vice versa. While mutual information can detect linear and nonlinear statistical dependencies between two time series. It has a maximum value when the two time series contain the same information and has the value of zero if one system is completely independent of the other.

In this study, we use these two methods to examine the changes of the extracted features with BP during an excitation of vasodilation induced by administrating arterial drugs.

## III. RESULTS

Data obtained in previous studies using correlation coefficient indicated that PTT has different degree of correlation with BP, varying from -0.49 to -0.69 at rest [11]. According to Sawada et al [12], there is no consistent tendencies between PTT and BP under various conditions, and the low correlation is insufficient to warrant the use of PTT as alternative index of BP. In our study, we investigated 222 indicators including PTT, with their relationship with BP being evaluated with CC, CSE, and MI. Results were compared with CC measured at central aortic and peripheral site with CC measured when there is BP changes induced by drugs. As can be seen in Fig. 2, the correlation coefficients between extracted features and BP varied at different conditions. It is obvious that their correlation relationship strengthened when the nitroglycerin was administrated. Whereas under rest state with intra-arterial measured in the aortic artery and peripheral radial artery, the correlations were comparable. Regarding PTT, different PTT calculated from different feature points correlated differently with different BP component, and at different scenarios. However, except for $PTT_5$ correlates well with SBP and PP, none of other PTTs has consistent correlation with BP.

It can be observed that the best correlated features for different BP components were different, although the pattern for different features were quite similar. More specifically, the pulse width at half amplitude of PPG had the highest correlation with SBP and MBP, the amplitude between dPPG_peak and sdPPG_b for DBP, and PTT calculated from R peak of ECG and sdPPG_a for PP.

The averaged CSE, MI as well as CC between these features and BP, with SBP as a representative example, is showed in Fig. 3. In general, the higher the value of the absolute CC, the higher the value of the MI, and the lower the CSE. As CSE and MI could provide the measure of relationship with a value within the range of 0 and 1, it is easier to assess association of two variables than using correlation coefficient. With the integration of these three measures, it is found that the indicator – pulse width at half amplitude of PPG has the highest CC, lowest CSE and highest MI value for SBP.

With comprehensive evaluation of these three methods, five of the best features were chosen for SBP, DBP, MBP, and PP, which are listed in Table II. The pulse width at half of the pulse amplitude is the most significant biomarker for SBP and MBP, the intensity and amplitude features for DBP, and PTT feature for PP.

The correlation coefficients of beat-to-beat BP with all the possible features of two subjects were illustrated in Fig. 4. We may observe that some of the features, e.g. PW and AM had a consistently high correlation with SBP each individual; whereas some features have very different patterns, either with varying amplitude of correlation or with reverse phase. These indicate that for different subject, features have different relationship with BP. It is very important to identify the features that have consistently strong association with BP through all different condition for each specific subject, ensuring the precise estimation of BP for individual and across all population.

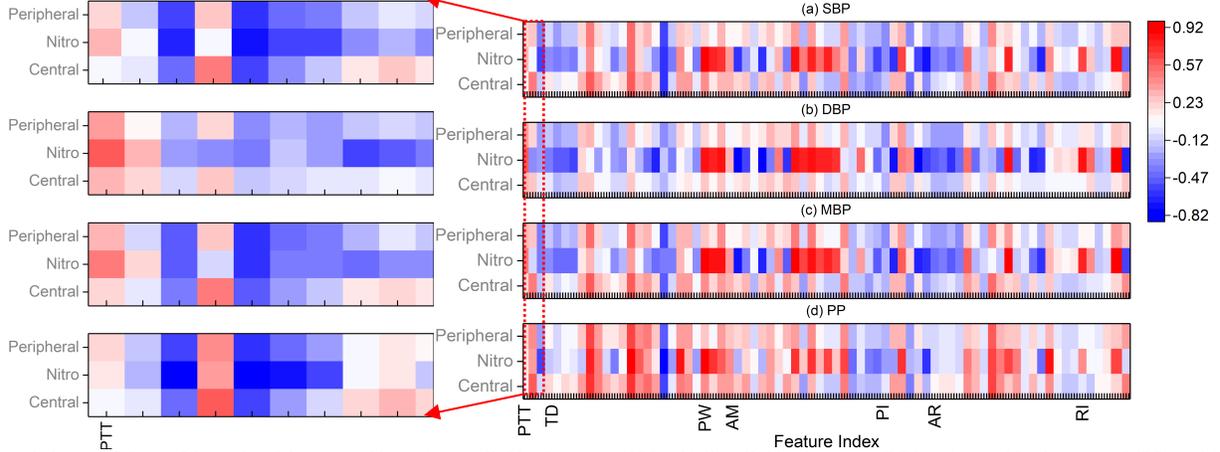

Fig. 2. Correlation coefficients of the explored features and intra- (a) systolic blood pressure (SBP), (b) diastolic blood pressure (DBP), (c) mean blood pressure (MBP), and (d) pulse pressure (PP) at central aortic at rest, while administering nitroglycerin, and at peripheral radial artery.

TABLE II. The Five Best Features for Each of BP Component

|  | SBP | DBP | MBP | PP |
| --- | --- | --- | --- | --- |
| Top Feature 1 | Pulse Width 50% | Amplitude between dPPG_max and sdPPG_b | Pulse Width 50% | Time interval from R_peak to sdPPG_a |
| Top Feature 2 | Pulse Width with dPPG_max | Intensity of sdPPG_b | Pulse Width with dPPG_max | Time interval from R_peak to dPPG peak |
| Top Feature 3 | Pulse Width 60% | (c+d-b)/a | (b-c-d)/a | Time duration between dPPG peak and sdPPG_c |
| Top Feature 4 | Amplitude between PPG valley and sdPPG_d | (b-c-d)/a | (c+d-b)/a | Pulse Width 50% |
| Top Feature 5 | (c+d-b)/a | Amplitude between PPG valley and sdPPG_b | Amplitude between sdPPG_b and sdPPG_d | Pulse Width with dPPG_max |

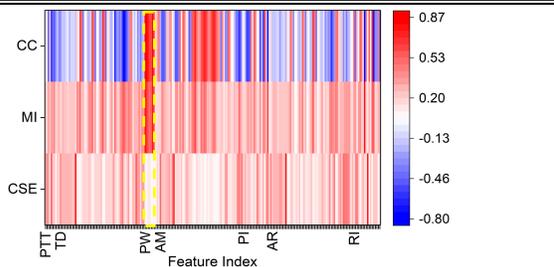

Fig. 3. The CSE, MI and CC of the explored features and SBP with the administration of nitroglycerin.

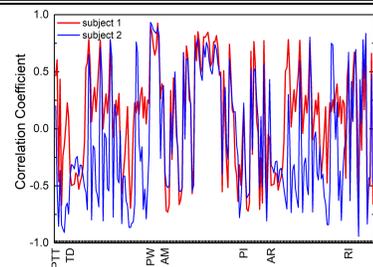

Fig. 4. CC of SBP and all the possible features for two subjects under the administration of nitroglycerin.

## IV. Discussion and Conclusion

Cuff-free continuous BP measurement within a clinically acceptable margin of error remains a challenge. On the one hand, the pattern of BP changes varies from time to time for one individual. On the other hand, the interaction between BP and the cardiovascular system varies from person to person. These are the major reasons for the large variance of error when we validate the cuffless BP estimation method following on specific standard, e.g. Association for the Advancement of Medical Instrumentation (AAMI), and IEEE Std 1708-2014 – IEEE Standard for Wearable Cuffless Blood Pressure Measuring Devices. There is therefore a pressing need to identify the optimal indicators/biomarkers and the model to calibrate the indicators to BP that would be able to reflect the BP changes constantly no matter what the condition and the subject is.

With the wide application of artificial intelligence (AI) in medical problem and healthcare, there has been some attempts towards using AI as a powerful computation tool to investigate the noninvasive cuffless BP estimation/prediction. Shukla [13], for example, reports that BP estimation with 10 features extracted from PPG using Multi Task Gaussian Processes and Artificial Neural Network achieved an improved accuracy compared with existing method. However, these studies have either employed incomplete features that extracted from PPG signal or without causal inference analysis of the features and target BP. In this study, we obtained 222 features from PPG signals and examined its relationship with BP under static and drug-induced conditions.

We found that each BP component, *i.e.*, SBP, DBP, MBP, and PP, was associated with different features. This finding is in line with the regulation mechanisms of BP. In addition, PTT was not the most relevant indicator of BP changes, except for PP. Rather, the Full Width with Half Maximum (FWHM) was the best features for SBP and MBP. Previous study demonstrates that the FWHM can represent the systemic peripheral resistance, and this study therefore demonstrates that the multiple features are potential to depict various determinants of BP.

Most notably, this study attempted to analyze the coupling relationship between the features and BP not only in terms of correlation coefficient but also with cross sample entropy and mutual information. Our results provide insights into the feature selection for the next step of modeling.

However, some limitations are worth noting. This study is a proof of concept for multiple feature exploration. Although the hypothesis was supported by case study, the sample size was not statistically significant. The other limitation is that the collection of the data was not long term, which made the features mainly in time-domain, rather than frequency-domain. However, frequency-domain as well as some nonlinear features may also be meaningful. Future work should therefore include larger sample size to evaluate the efficiency of this work and to apply it for BP estimation with all-around extracted features and the consideration to integrate physiological theory guided model and the data-driven model.

Acknowledgment

The authors would like to thank Dr. Jing Liu and Peng Su from the Department of Electronic Engineering of The Chinese University of Hong Kong, and the nurses of Princes Wales Hospital, Hung-Sing Lai and Vincent Tsz for their assistance to conduct the clinical experiment.